\pdfoutput=1
\documentclass[acmtog]{acmart}

\AtBeginDocument{%
  \fancypagestyle{plain}{%
    \fancyhf{} %
    \fancyfoot[L]{ACM SIGENERGY Energy Informatics Review}%
    \fancyfoot[R]{Volume 3 Issue 3, October 2023}}
  \providecommand\BibTeX{{%
    \normalfont B\kern-0.5em{\scshape i\kern-0.25em b}\kern-0.8em\TeX}}}
    
\let\oldmaketitle\maketitle
\renewcommand{\maketitle}{%
  \oldmaketitle%
  \thispagestyle{plain}%
  \pagestyle{plain}}

\usepackage{tikz}
\usepackage{amsmath}
\usepackage{multirow}
\usepackage{xurl}
\usepackage{enumitem}
\usepackage{tcolorbox}
\usepackage{amsmath,amsfonts}
\usepackage{graphicx}
\usepackage{textcomp}
\usepackage{xcolor}
\usepackage{balance}
\hypersetup{
     colorlinks=true,
     linkcolor=blue,
     filecolor=blue,
     citecolor =blue,      
     urlcolor=blue,
     }

\usepackage{filecontents}    
{ \vspace{-0.15cm}%
    \small\noindent{\bfseries Availability of Data and Material:}\par%
    \noindent\ignorespaces}%
{ \par\noindent%
\ignorespacesafterend }%

\setcopyright{acmcopyright}
\copyrightyear{2023}
\acmYear{2023}
\acmDOI{XXXXXXX.XXXXXXX}

\begin{document}

\title{The Dirty Secret of SSDs: Embodied Carbon}

\author{Swamit Tannu}
\email{swamit@cs.wisc.edu}
\orcid{0000-0003-4479-7413}
\affiliation{%
  \institution{University of Wisconsin, Madison}
  \streetaddress{1210 W. Dayton Street}
  \city{Madison}
  \state{Wisconsin}
  \country{USA}
  \postcode{53706-1613}
}

\author{Prashant J. Nair}
\email{prashantnair@ece.ubc.ca}
\orcid{0000-0002-1732-4314}
\affiliation{%
  \institution{University of British Columbia}
  \streetaddress{2332 Main Mall}
  \city{Vancouver}
  \state{British Columbia}
  \country{Canada}
  \postcode{V6T 1Z4}
}

\renewcommand{\shortauthors}{Tannu, S. and Nair, P.J.}

\begin{abstract}
Scalable Solid-State Drives (SSDs) have ushered in a transformative era in data storage and accessibility, spanning both data centers and portable devices. However, the strides made in scaling this technology can bear significant environmental consequences. On a global scale, a notable portion of semiconductor manufacturing relies on electricity derived from coal and natural gas sources. A striking example of this is the manufacturing process for a single Gigabyte of Flash memory, which emits approximately 0.16 Kg of CO$_{2}$ -- a considerable fraction of the total carbon emissions attributed to the system. Remarkably, the manufacturing of storage devices alone contributed to an estimated 20 million metric tonnes of CO$_{2}$ emissions in the year 2021.

In light of these environmental concerns, this paper delves into an analysis of the sustainability trade-offs inherent in Solid-State Drives (SSDs) when compared to traditional Hard Disk Drives (HDDs). Moreover, this study proposes methodologies to gauge the embodied carbon costs associated with storage systems effectively. The research encompasses four key strategies to enhance the sustainability of storage systems.

Firstly, the paper offers insightful guidance for selecting the most suitable storage medium, be it SSDs or HDDs, considering the broader ecological impact. Secondly, the paper advocates for implementing techniques that extend the lifespan of SSDs, thereby mitigating premature replacements and their attendant environmental toll. Thirdly, the paper emphasizes the need for efficient recycling and reuse of high-density multi-level cell-based SSDs, underscoring the significance of minimizing electronic waste.

Lastly, for handheld devices, the paper underscores the potential of harnessing the elasticity offered by cloud storage solutions as a means to curtail the ecological repercussions of localized data storage. In summation, this study critically addresses the embodied carbon issues associated with SSDs, comparing them with HDDs, and proposes a comprehensive framework of strategies to enhance the sustainability of storage systems.
\end{abstract}



\maketitle

\section{Introduction}
The carbon footprint of computing systems is multifaceted, encompassing emissions throughout their lifecycle, from manufacturing and operation to transportation and recycling. Of particular concern are the billions of hand-held devices, including smartphones, tablets, and web services, that have become integral to modern life. This proliferation of devices has contributed significantly to global warming, with the current combined carbon emissions from computing and networking devices already accounting for approximately 2\% of the total carbon emissions~\cite{ICT1, ICT2}. Projections suggest that this percentage could double within the coming decade, underscoring the urgency of addressing these emissions. As digital data creation and consumption continue to surge worldwide, a comprehensive understanding of the carbon emissions from personal devices, data centers, and networking infrastructure—collectively known as the Information and Communication Technologies (ICT) sector—becomes paramount.

The majority of carbon emissions within the ICT sector stems from the utilization of ``conventional" electricity sources~\cite{greenelec}, which play a pivotal role in both the manufacturing and operational phases of computing systems~\cite{gupta2022chasing}. The energy-intensive tasks of running and cooling computing and networking hardware translate to substantial electricity consumption. When this electricity is sourced from carbon-intensive fuels such as coal, natural gas, and crude oil, the resulting emissions contribute significantly to global warming. Conversely, electricity generated from renewable sources—such as wind, solar, nuclear, and hydroelectric—exhibits a considerably smaller Global Warming Potential (GWP). Nonetheless, a prevailing challenge persists: whether in the context of hand-held devices or server nodes, the manufacturing and operation of hardware invariably demand substantial electricity, often originating from carbon-intensive conventional sources.

\begin{figure}[b]
	\centering
	\includegraphics[width=\columnwidth]{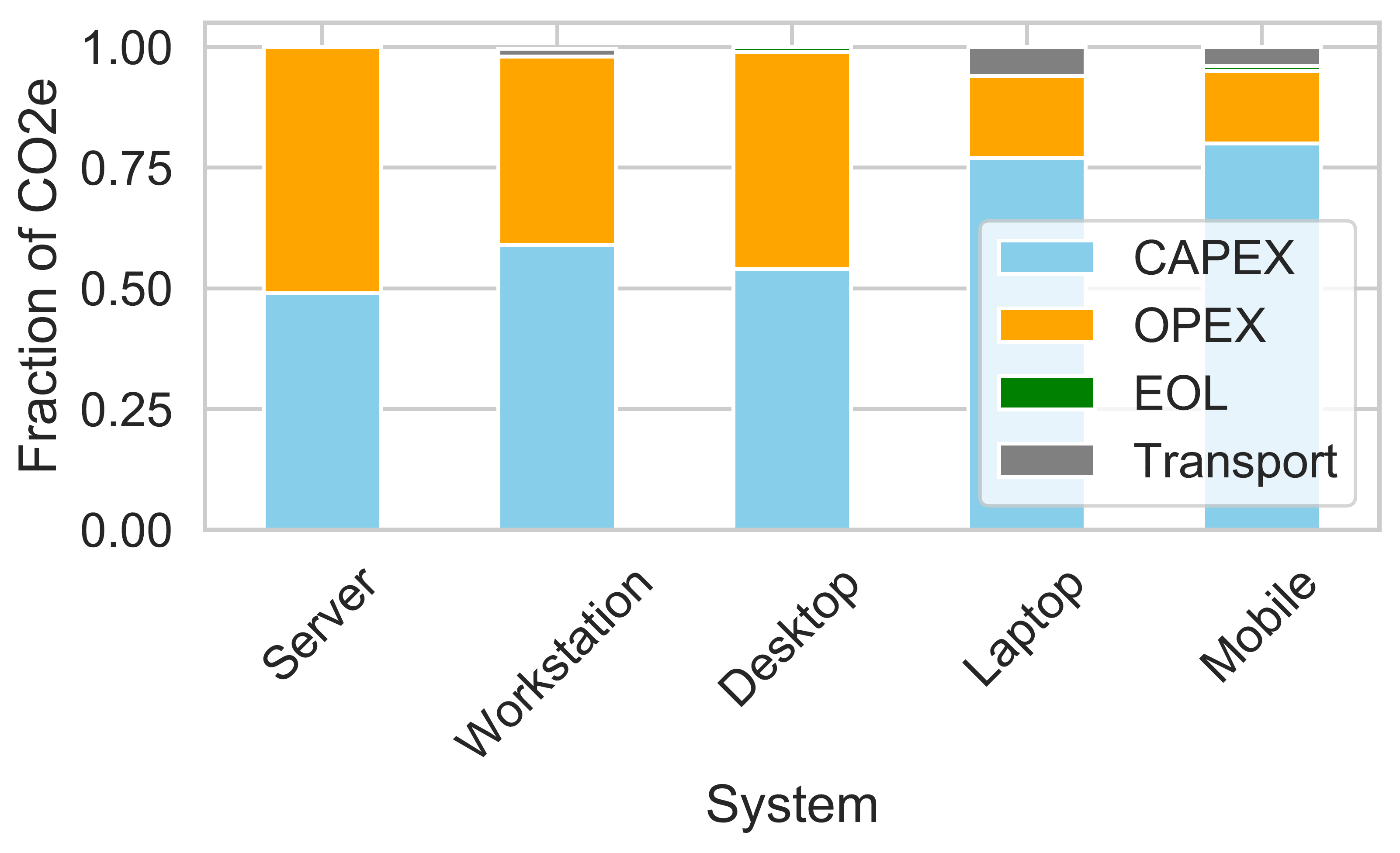}
	\caption{Breakdown of CO2e in Manufacturing (CAPEX) Operations (OPEX), Transport, and End of Life (EOL) phases.}
	\label{fig:emissions}
\end{figure}   

\noindent \textbf{Estimating Global Warming Potential:} The Global Warming Potential (GWP) is a critical metric that gauges the quantity of emitted CO$_2$, or its equivalent in CO2e measured in kilograms~\footnote{All greenhouse gas emissions are standardized against the global warming potential of CO$_2$. For instance, Methane possesses a 25$\times$ higher global warming potential than CO$_2$, meaning that 1 kg of Methane emission is equivalent to 25 kg of CO2e.}. In this paper, we consistently employ CO2e as the metric for quantifying the carbon footprint of various processes. For instance, the production and assembly of all components in an iPhone 13 release a total of 78 kg CO2e~\cite{iphone13LCA}. This is regarded as the Capital Expenditure (CAPEX) phase. As depicted in Figure~\ref{fig:emissions}, the emissions associated with iPhone operation over a five-year span constitute a mere 15\% in contrast to the dominant 80\% attributed to manufacturing or CAPEX phase. Similarly, the bulk of carbon emissions for laptop-type systems arises from the manufacturing or CAPEX phase. In comparison, CO$_2$ emissions from server systems are primarily concentrated in the Operational Expenditure (OPEX) phase.

\vspace{0.05in}

\begin{figure}[t]
	\centering
	\includegraphics[width=\columnwidth]{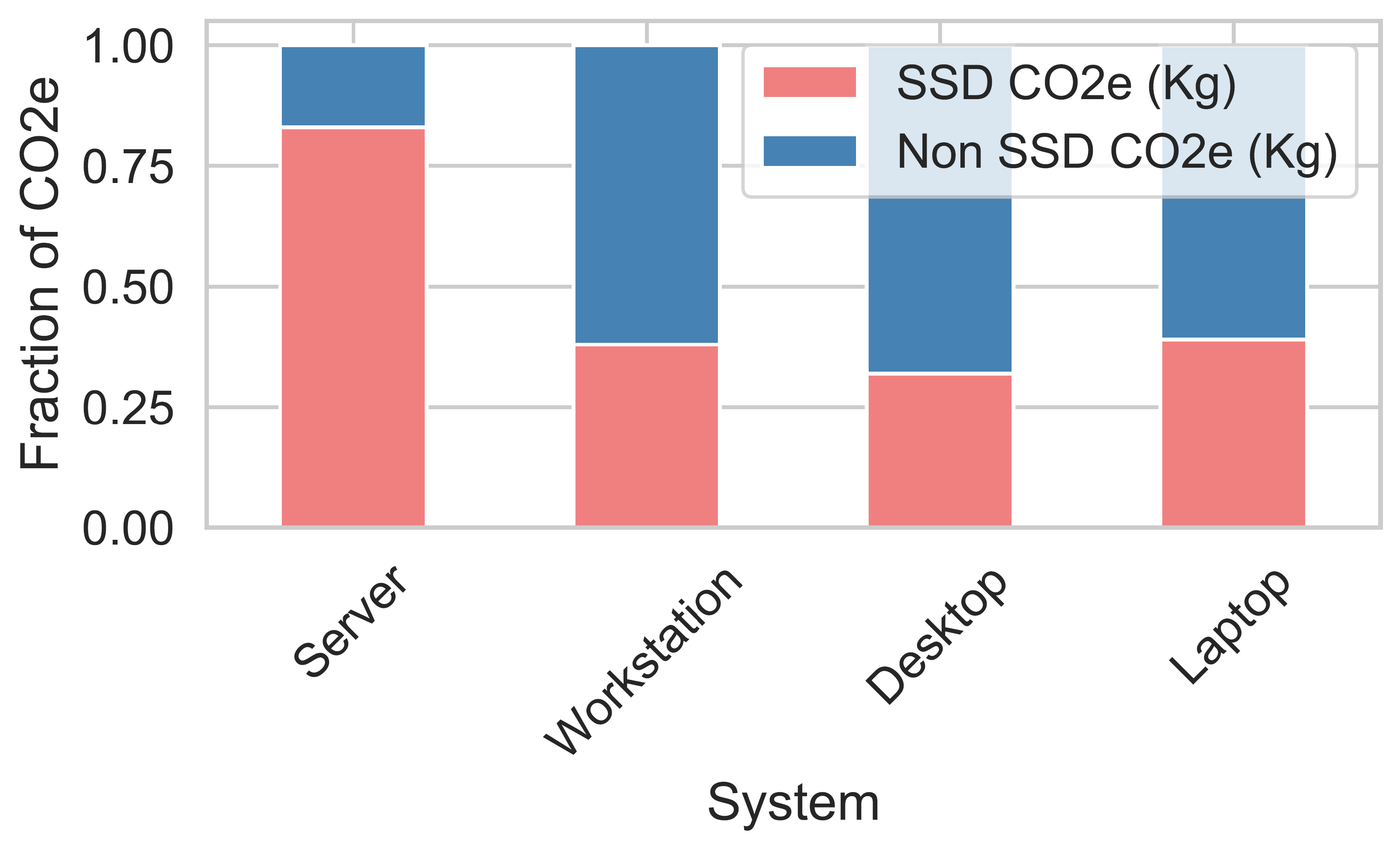}
	\caption{Contributions of SSDs to Embodied CO2e in Four Computing Systems - Data sourced from~\cite{LCADellR740,Dell_LCA,applemac73:online,macminie74:online}.}
	\label{fig:ssdCO$_2$}
\end{figure}

\noindent \textbf{Identifying CO2e Hotspots in Systems:} Memory and storage represent foundational elements across all computing platforms. Regrettably, the manufacturing procedures employed for modern Flash and DRAM devices are markedly energy-intensive. Consequently, a significant proportion of CAPEX CO2e can be attributed to Solid State Drives, as illustrated in Figure~\ref{fig:ssdCO$_2$}. This paper embarks on an in-depth analysis of a Life Cycle Assessment (LCA) report dataset, meticulously quantifying the CO2e associated with both Solid State Drives (SSDs) and Hard Disk Drives (HDDs). Our investigation reveals that, on average, SSDs significantly contribute to embodied carbon costs. They have about 8$\times$ higher embodied cost than Hard Disk Drives (HDDs) with an identical capacity.

\vspace{0.05in}

\noindent Moving forward, it is essential to answer the following questions:
\begin{enumerate}[leftmargin=*]
\setlength\itemsep{0em}
    \item What constitutes the comprehensive carbon footprint of SSDs?
    \item How does the overall sustainability of HDDs compare with that of SSDs?
    \item Which methodologies can be formulated to estimate the embodied carbon cost of storage systems accurately?
    \item What strategies can we employ to facilitate the emergence of sustainable storage systems?
\end{enumerate}

\section{Problem: Embodied Carbon Emissions}
We begin by dissecting the embodied carbon emissions of a typical desktop system and conducting an analysis of CO2e for both SSDs and HDDs. Additionally, we delve into the exploration of straightforward yet impactful metrics designed to quantify the CO2e associated with storage systems.

\subsection{System-level Breakdown of Embodied Cost}

\begin{figure}[t]
	\centering
	\includegraphics[width=0.86\columnwidth]{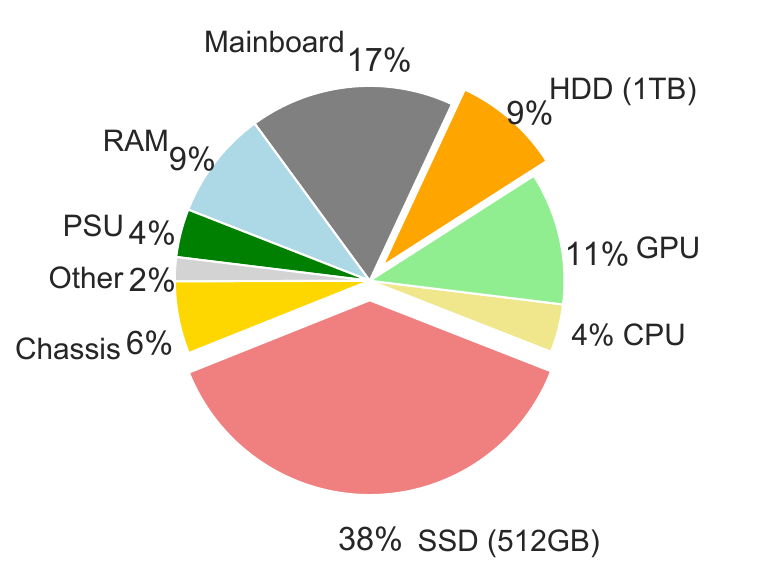}
	\caption{Distribution of embodied CO2e across different components of a Desktop system, data derived from~\cite{LCA_Fujutsu}.}
	\label{fig:system_split}
\end{figure}

Figure~\ref{fig:system_split} shows the breakdown of embodied CO2e, sourced from the Life Cycle Analysis (LCA) report detailed in~\cite{LCA_Fujutsu}. The desktop configuration encompasses a four-core CPU, 8 GB DRAM, a 1TB HDD, and a 512GB SSD. This system's cumulative CO2e reaches 706 kg CO$_2$, comprising OPEX carbon emissions of 278 kg CO$_2$ over a five-year lifespan and CAPEX carbon emissions totalling 473 kg CO$_2$. The LCA methodology, as discussed in~\cite{LCAwiki}, provides an all-encompassing approach that traces carbon emissions from inception to conclusion. The assessment encompasses carbon emissions originating from material extraction during the mining, refining, and transportation phases. Subsequently, it models carbon emissions entailed in the manufacturing of components such as PCBs, Integrated Circuits, and Chassis. Lastly, the LCA computes CO$_2$ emissions associated with operational, transportation, and recycling phases. Notably, LCA software employs intricate models that consider material specifics, process technologies, packaging methods, yield, and geographical manufacturing nuances.

In the case of the embodied CO2e within the Fujitsu workstation depicted in Figure~\ref{fig:system_split}, the bulk of embodied carbon emissions is attributed to semiconductor components—namely SSDs, DRAM, HDDs, and CPUs. This outcome is a direct consequence of the intricate and energy-intensive nature characterizing the manufacturing of semiconductor components. Within this component spectrum, SSDs claim the most substantial footprint due to their incorporation of multiple Flash and DRAM chips, often containing numerous silicon dies within a single package. Adding to the challenge, the Flash and DRAM fabrication centers have limited renewable electricity supply, thereby compelling these facilities to use electricity sourced from carbon-intensive origins.

\begin{tcolorbox}
\textbf{Key Observation 1:} A substantial proportion of carbon dioxide (CO$_2$) emissions can be attributed to the manufacturing processes of semiconductors. This proportion shows a notable increase in correlation with technology scaling.

\end{tcolorbox}

\subsection{Embodied Carbon Cost of SSDs}

We conducted an analysis encompassing 94 Life Cycle Assessment (LCA) reports, which collectively quantify the embodied cost of SSDs. Owing to the scarcity of direct and up-to-date LCA studies focused specifically on SSDs. We compiled a dataset comprising LCA reports pertaining to Server, Workstation, Desktop, Laptop, and Chromebook products, all of which feature SSDs~\cite{HP_LCA, Dell_LCA, Apple_LCA, LCA_Seagate, LCADellR740}.

\begin{figure}[t]
	\centering
	\includegraphics[width=\columnwidth]{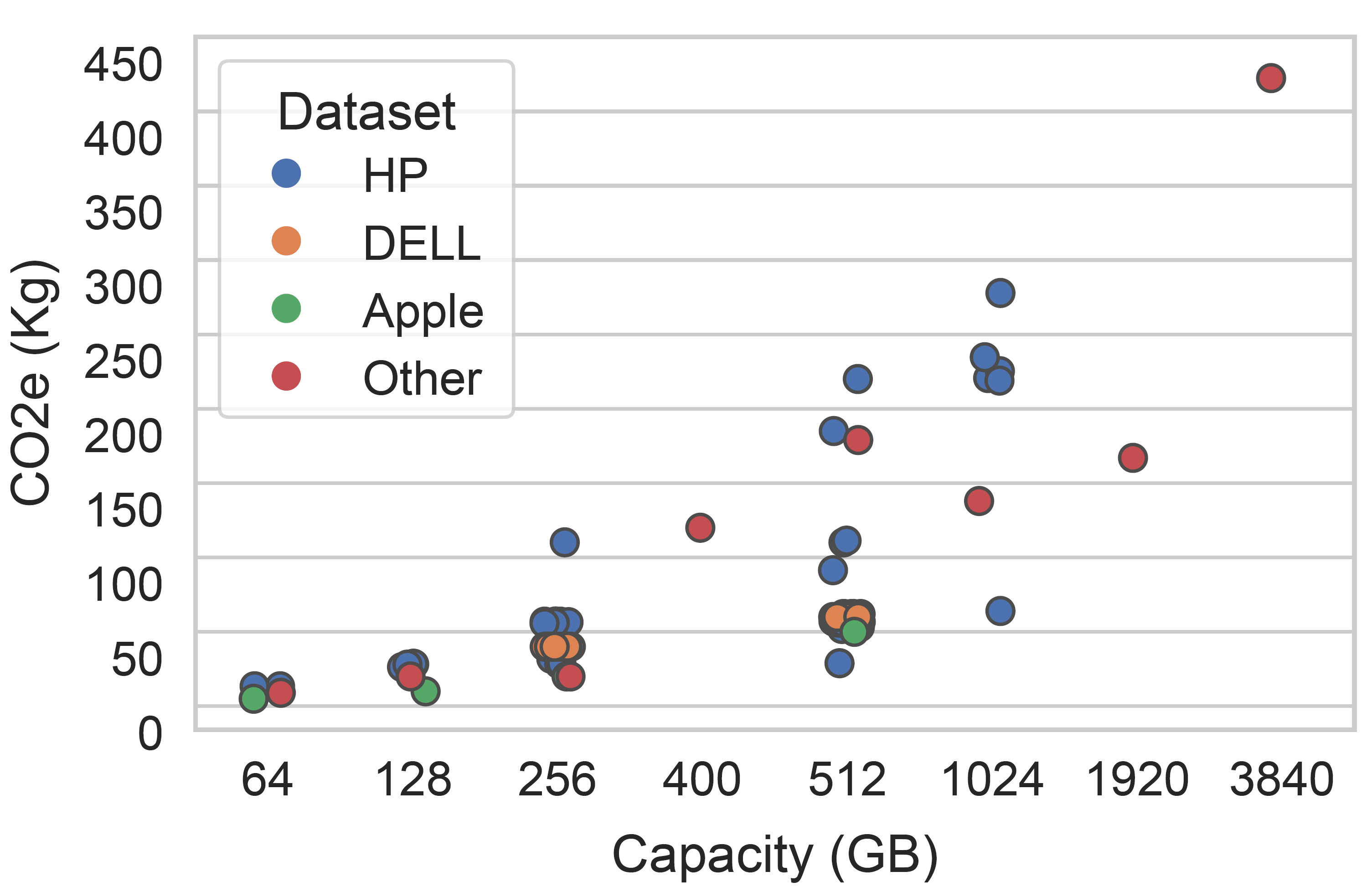}
	\caption{Carbon emissions for manufacturing 94 Solid State Drives (SSDs). Data based on Life Cycle Analysis (LCA) reports published by eight vendors.}
	\label{fig:lca_ssd}
    \vspace{-0.2in}
\end{figure}

Figure~\ref{fig:lca_ssd} visually encapsulates the CO2e data gleaned from these five distinct datasets. Each dataset encompasses a diverse array of devices, reflecting variations in capacity, technological nodes, and device types. Notably, these LCA reports are often generated by the vendors responsible for assembling the computing systems, such as Dell, HP, Apple, Fujitsu, etc. This data is organized as per the increasing capacity of SSDs. Notably, it is important to acknowledge that identical SSD capacities can yield substantially different embodied costs. This is because the amount of CO2e is contingent upon several influencing factors.

\begin{figure}[t]
	\centering
	\includegraphics[width=\columnwidth]{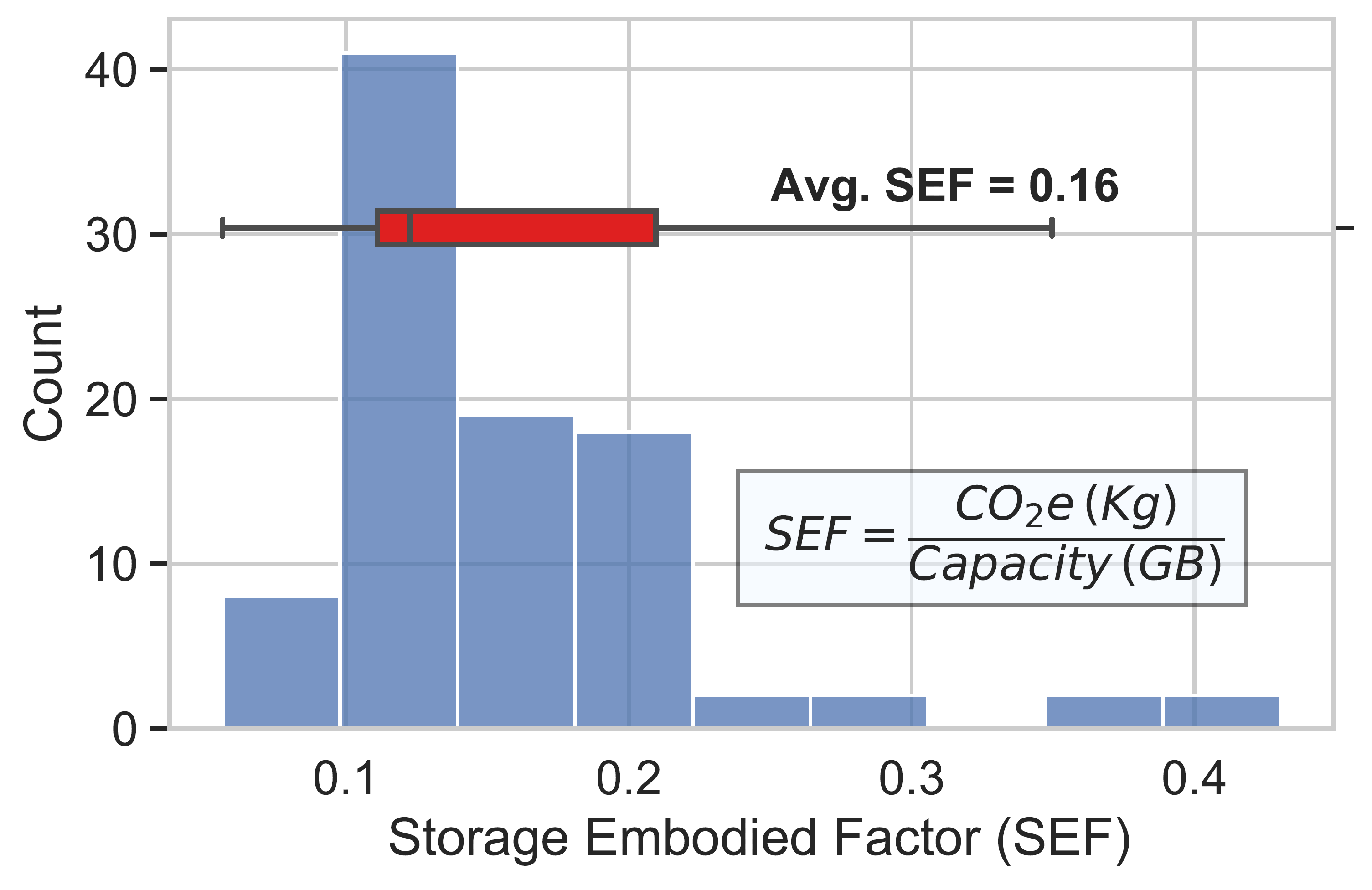}
	\caption{Distribution of estimated storage embodied factor for 94 Solid State Disks (SSDs).}
	\label{fig:lcapergb}
\end{figure}

While such disparities exist, a consistent trend emerges upon analysis. On average, the CO2e exhibits a linear growth trajectory in proportion to expanding SSD capacity. This observation, obtained from a diverse array of LCAs, provides a valuable perspective on the evolving embodied cost of flash storage. Figure~\ref{fig:lcapergb} delineates the distribution of CO2e per gigabyte of flash storage and paints a more nuanced picture of the carbon footprint associated with SSDs.

To facilitate a quantitative assessment of storage devices' embodied cost, which is crucial for the development of sustainable storage architecture, we introduce the concept of the Storage Embodied Factor (SEF). As illustrated in Figure~\ref{fig:lcapergb}, SEF is a ratio of CO2e and the capacity of the storage medium. Our evaluations yield an average SEF value of 0.16 kg CO2e per gigabyte (GB) for SSDs. We validate our result using four-fold cross-validation. 

\subsection{Embodied Carbon Cost of HDDs}

A noteworthy distinction emerges between SSDs and HDDs, with SSDs demonstrating remarkable energy efficiency due to their lack of moving parts. This pivotal difference substantially curtails both idling and active power consumption. Furthermore, SSDs boast enhanced attributes like higher bandwidth and reduced latency, contributing to an overall improvement in system energy efficiency.

In contrast, HDDs exhibit a bulkier design and necessitate a greater quantity of materials during the manufacturing process. Logically, one might anticipate that HDDs would possess a higher embodied CO2e when compared to SSDs. However, our analysis of previously published Life Cycle Assessment (LCA) reports presents a counterintuitive revelation.

\begin{figure}[b]
	\centering
	\includegraphics[width=\columnwidth]{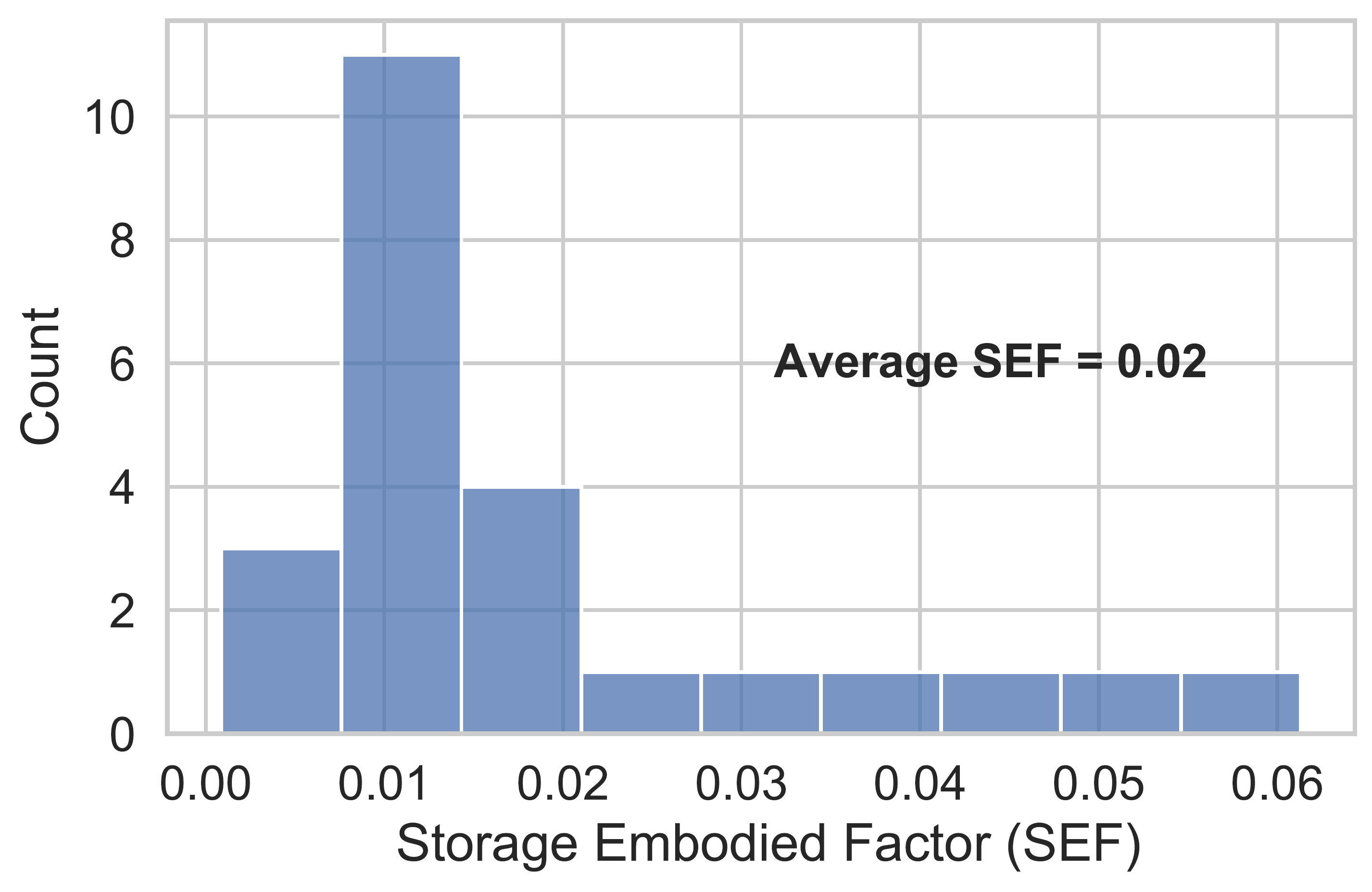}
	\caption{Distribution of estimated storage embodied factor for 24 Hard Drive Disks (HDDs).}
	\label{fig:hdd}
\end{figure}  

In our comprehensive examination, we compiled data from 24 LCA reports, focusing on the SEF for HDDs produced by four distinct vendors. These reports encompassed a range of capacities spanning from 512GB to 6TB. Figure~\ref{fig:hdd} provides a visual representation of the distribution of SEF alongside the average SEF values.

\begin{tcolorbox}
\textbf{Key Observation 2:} When juxtaposed against SSDs, the embodied carbon cost of HDDs proves to be at least an order of magnitude lower. This intriguing outcome challenges conventional expectations. It underscores the importance of a nuanced understanding of the dynamics of embodied carbon costs for different storage technologies.
\end{tcolorbox}  

\subsection{Impact of Technology Scaling on CO2e}

\begin{figure}[b]
	\centering
	\includegraphics[width=1\columnwidth]{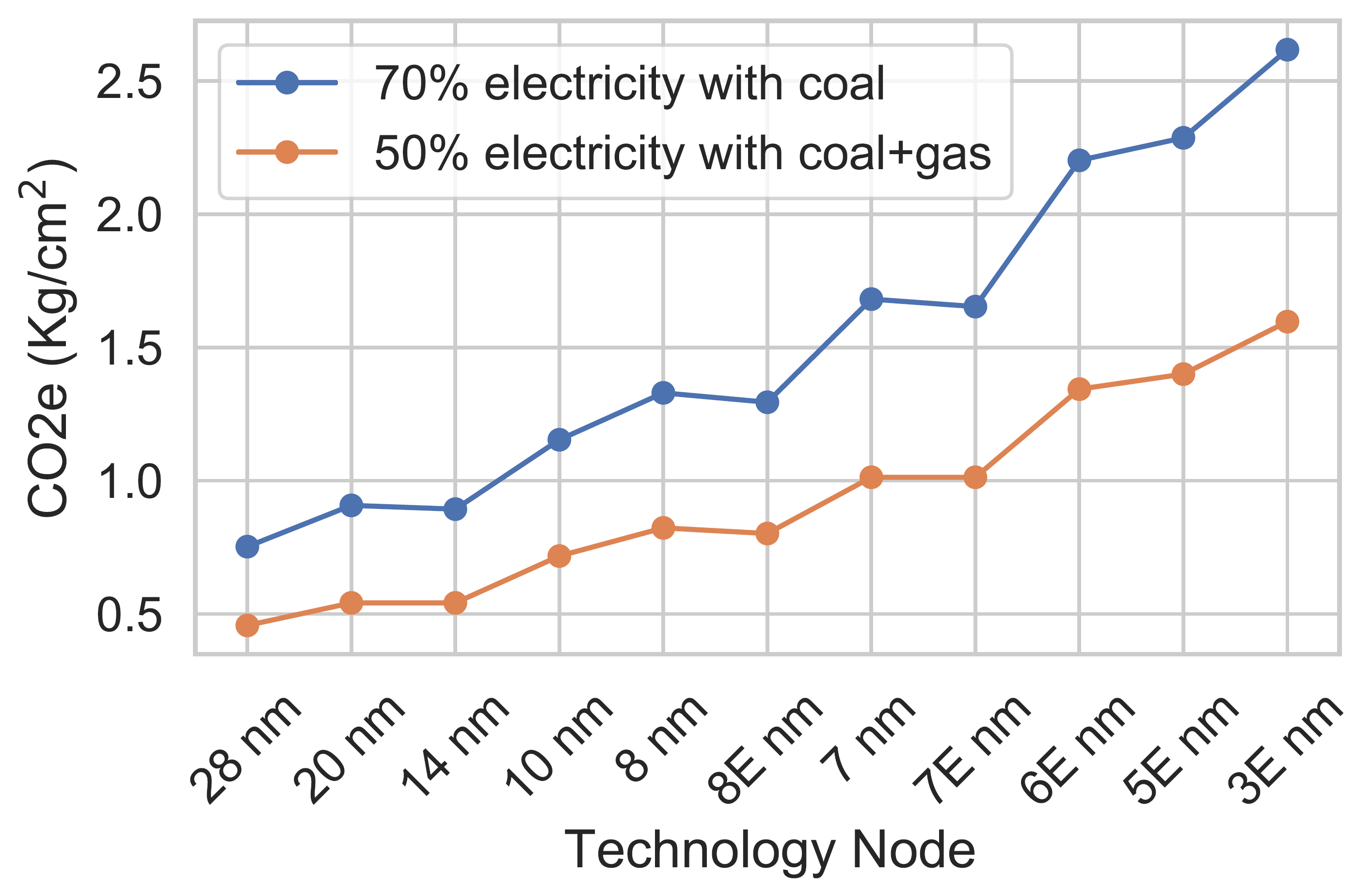}
	\caption{Carbon emissions for manufacturing cm$^2$ of silicon die for different technology nodes for two electricity generation schemes.}
	\label{fig:technology}
\end{figure}

Efforts to enhance Flash storage density revolve around packing a greater number of transistors within a fixed chip area. This goal is achieved through transistor scaling and the innovative technique of 3D die stacking. Reducing the size of transistors allows for a higher transistor count within the chip, while layering multiple wafer dies on top of each other contributes to expanded capacity. Regrettably, these strategies demand intricate and energy-intensive manufacturing processes. Furthermore, as transistor (technology) scaling advances, the intricacies of chip manufacturing escalate, resulting in a rapid increase in carbon emissions associated with silicon die fabrication~\cite{imec}.

As shown in Figure~\ref{fig:technology}, the carbon emissions stemming from the fabrication of a logic die exhibit variations across different technology nodes. Notably, the reduction in transistor feature sizes correlates with a heightened number of fabrication steps and their associated energy intensity. This, in turn, contributes to elevated carbon emissions. A pivotal reason for these substantial emissions lies in the scarcity of renewable energy sources at the primary locations of semiconductor manufacturing. This inadequacy drives the reliance on electricity generated by coal and natural gas plants in most semiconductor fabrication facilities.

The data in Figure~\ref{fig:technology} originates from a recent study~\cite{imec}. This study provides a detailed analysis of the impact of using carbon-intensive sources for chip manufacturing. Presently, only a modest fraction of the electricity harnessed in semiconductor manufacturing stems from renewable sources~\cite{cnbc}. While semiconductor manufacturers steadfastly pursue carbon neutrality, transitioning from coal to renewable energy sources poses formidable challenges and continues to unfold slowly.




\section{Strategies to Reduce Storage CO2e}

We put forth three pivotal strategies designed to curtail the embodied carbon costs associated with storage devices effectively. These strategies draw upon considerations of selecting the optimal storage medium, enhancing resilience and endurance frameworks, advocating for recycling and reuse practices, and harnessing the inherent elasticity inherent within cloud storage solutions.

\subsection{Selecting the Right Storage Medium}

At a surface level, the energy-proportional nature of Flash-based storage sets it apart from Hard Disk Drives (HDDs), which contend with significant idle power consumption~\cite{energyprop}. Common wisdom dictates that transitioning from HDDs to SSDs would confer energy savings and foster a greener storage landscape~\cite{IntelSSD}. Yet, we assert that a comprehensive evaluation mandates consideration of both Operational Expenditure (OPEX) CO2e and Embodied Cost when determining the ideal storage architecture.

To illustrate this perspective, we embark on an assessment juxtaposing the Capital Expenditure (CAPEX) and OPEX CO2e associated with 1TB SSDs and HDDs. Our approach encompasses leveraging the average Storage Embodied Factor (SEF) data alongside the mean power consumption of both HDD and SSD devices. Considering a workload involving 20\% active and 80\% idle cycles, Table~\ref{tab:ssdhdd} furnishes an estimation of energy consumed over five and ten years for the two storage mediums.

In our model, an average HDD power consumption of 4.2 W is assumed, in contrast to the 1.3 W consumption by SSDs. OPEX CO2e is determined through total energy usage, employing an emission factor delineated by the Environmental Protection Agency (EPA)\footnote{The emission factor of 0.7 Kg CO2 per kWh is grounded in the U.S. national weighted average CO2 marginal emission rate based on 2019 data~\cite{EPA_Cal}}. The CAPEX calculation hinges on the average SEF for both SSDs and HDDs. Additionally, we account for a CAPEX upgrade cost given the five-year lifespan of both storage types, thus extending the analysis to a decade.

It emerges from Table~\ref{tab:ssdhdd} that, intriguingly, the overall CO2e for HDDs notably trails that of SSDs. This apparent contradiction is primarily attributed to the relatively lower embodied cost characterizing HDDs. However, this first-order initial assessment does not account for SSDs' potential impact on broader performance dynamics, power utilization, and overall energy consumption.

\begin{table}[h!]
\centering
\caption{Emissions of SSD and HDD over 5-year and 10-year lifetimes.}
\setlength{\tabcolsep}{0.1cm} 
\renewcommand{\arraystretch}{1.25}
\resizebox{1.0\columnwidth}{!}{
\begin{tabular}{ | c || c | c || c | c || c | c || c | c |}
\hline
    \multirow{2}{*}{Storage} &  \multicolumn{2}{c||}{Energy} &  \multicolumn{2}{c||}{OPEX} & \multicolumn{2}{c||}{CAPEX}  &  \multicolumn{2}{c|}{Total} \\
     & \multicolumn{2}{c||}{(KWh)} & \multicolumn{2}{c||}{CO2e (Kg)}  & \multicolumn{2}{c||}{CO2e (Kg)}  &  \multicolumn{2}{c|}{CO2e (Kg)} \\
    \cline{1-9}
    Exp. Life & 5yr  & 10yr & 5yr & 10yr &  5yr & 10yr & 5yr & 10yr  \\
    \hline \hline
    
    HDD (1TB) & 183.9  & 367.9  &  79.6  & 159 & 20 & 40 & 99.6  & 199  \\
    \hline
    SSD (1TB) & 56.9 & 113.8 & 24.6 & 49.2 & 160 & 320  & 184 & 369.2 \\
    \hline
        
\end{tabular}}
\label{tab:ssdhdd}

\end{table}

\subsection{Extending the Lifetime of SSDs}
The embodied carbon cost can also be amortized by extending the lifetime of SSD devices. Here, we advocate using four approaches.

\subsubsection{Inter-Node Wear Leveling} 

Modern Flash-based SSDs exhibit limited endurance, capable of only 10,000 to 100,000 write cycles per cell~\cite{flashendurance}. To bolster endurance, contemporary SSDs employ table-based wear-levelling techniques~\cite{wearleveling, WLflash}. These strategies involve dynamic data redirection to locations with lower write intensity, ensuring even wear distribution across SSD cells. To enhance this technique, exploring wear-levelling strategies across multiple storage nodes would be invaluable. This exploration would delve into the trade-offs between elevated access latencies and the prolongation of Flash device lifespans.

\subsubsection{Intelligent Data Placement: SLC versus MLC}

The endurance of a flash device hinges on whether it utilizes single-level cells (SLC) or multi-level cells (MLC)\cite{hybridssd}. MLCs store multiple bits within each cell, augmenting storage capacity\cite{mlccharacterization}. However, MLC cell writes are inherently more taxing and diminish endurance significantly compared to SLCs. This disparity directly impacts flash device longevity and density. Looking ahead, reconsidering the distribution of user data within systems featuring both SLC and MLC flash devices can optimize their effective lifespans. Techniques like Zoned Namespaces (ZNS), tailored for performance isolation, can be fine-tuned to amplify SSD longevity and sustainability.

\subsubsection{Recycling and Reusing Flash Devices}

Recycling and reusing MLC devices as low-capacity SLC devices presents an intriguing prospect. This is primarily attributed to the rapid wear-out of MLC devices. Through transformation, these devices can potentially function as lower-capacity SLCs or contribute to hybrid storage mediums combining both SLC and MLC technologies. Although SLC devices may offer reduced capacity benefits from the standpoint of data centers or hand-held devices, repurposing these morphed SLC SSDs for other storage nodes (e.g., those managing data logging or storing dormant container images) holds promise.

\subsubsection{Efficient Error Correction Codes (ECC)} 

Enhancing the effective utilization of existing SSDs can be accomplished through the implementation of robust and efficient error-correcting codes (ECCs). Modern SSDs already integrate Low-Density Parity-Check Codes (LPDC), capable of rectifying multiple faulty bits within a data block~\cite{lpdc}. Custom-designed ECC codes, tailored to usage patterns, could bolster flash device lifetimes while incurring some capacity overhead. It's noteworthy that complex ECCs may introduce extended encoding and decoding latencies, subsequently elongating SSD access times~\cite{morphableeccdsn, flexecc}. This dynamic presents an intriguing avenue for exploration, underscoring the trade-offs between ECC complexity, access time, and flash device longevity.

\subsection{Leverage the Elasticity in Cloud Storage}

Unlike data centers, fabricating composable (or modular) hand-held systems presents notable challenges~\cite{composablesystems}. Recently, a renewed focus on repairing and recycling hand-held electronics has emerged~\cite{EUnews-rtr,apple-recycle}. While these strategies reduce effective CO2e overheads over the device's aggregate lifetime, they also require lifestyle adjustments for end users. Moreover, interconnecting components across disparate technology generations might prove less efficient. We can rethink data management across devices and the cloud to overcome these challenges.

Cloud storage harbours critical advantages, marked by scalability, security, composability, and heightened durability achieved through adept utilization of data redundancy~\cite{elasticcloud,tai2019s,alagappan2018fault,alagappan2018protocol,correia2012practical}. This infrastructure empowers manufacturers and service providers to determine which data remains locally stored on hand-held devices strategically. Additionally, they can formulate cloud storage pricing models that integrate the dimensions of embodied carbon costs, cloud latency, and network throughput into their computations. This approach ensures that cloud-based solutions effectively address the environmental implications associated with mobile device lifecycles.

\section{Summary}

Rapid technology advancements have enabled powerful flash-based SSDs to offer expansive storage, spanning from data centers to handheld gadgets. However, it's crucial to comprehend the environmental implications before fully embracing SSD-based systems. This paper sheds light on the embodied carbon expenses associated with SSDs during both their creation and operation. Our exploration demonstrates that SSDs typically account for a significant portion of total embodied carbon expenses.

In response, this paper compares the sustainability of SSDs and HDDs and presents practical methodologies to estimate carbon expenses in storage systems. It advocates for strategies such as selecting appropriate storage mediums, improving durability, recycling flash devices, and optimizing cloud storage for handheld devices. Our assertion is that the suggested transformations within storage architectures will not only guide the blueprint of long-term storage systems but also shape the design of countless devices. This, in turn, will ultimately lead to a reduction in their embodied carbon costs over the forthcoming decades. Our paper points the way toward an environmentally conscious future for storage systems, making a substantial impact on the design landscape of billions of devices.


\section*{Acknowledgements}
We thank the anonymous reviewers for their insightful suggestions and feedback. For this work, Prashant J. Nair was supported by the Natural Sciences and Engineering Research Council of Canada (NSERC) [funding reference number RGPIN-2019-05059].

\citestyle{acmauthoryear}
\bibliographystyle{ACM-Reference-Format}
\bibliography{ref}

\end{document}